\documentstyle[aps,prb,epsfig]{revtex}
\begin{document}
\author{V.~Ya.~Demikhovskii, A.~A.~Perov}
\address{Nizhny Novgorod State University, Gagarin ave., 23,
603950 Nizhny Novgorod, Russian Federation}
\title{Hall Conductance of a Two-Dimensional Electron Gas in Periodic
Lattice with Triangular Antidots}
\maketitle

\vspace{0.5cm}
\begin{abstract}
The topic of this contribution is the investigation of quantum states and
quantum Hall effect in electron gas subjected to a periodic potential of
the lateral lattice. The potential is formed by triangular quantum antidos
located on the sites of the square lattice. In a such system the inversion
center and the four-fold rotation symmetry are absent. The topological
invariants which characterize different magnetic subbands and their Hall
conductances are calculated. It is shown that the details of the antidot
geometry are crucial for the Hall conductance quantization rule. The critical
values of lattice parameters defining the shape of triangular antidots at which
the Hall conductance is changed drastically are determined. We demonstrate
that the quantum states and Hall conductance quantization law for the
triangular antidot lattice differ from the case of the square lattice
with cylindrical antidots. As an example, the Hall conductances of
magnetic subbands for different antidot geometries are calculated for the
case when the number of magnetic flux quanta per unit cell is equal to three.
\end{abstract}

\vspace{0.5cm}

\vspace{0.5cm}
\section{Introduction}
In a profound paper by Thouless, Kohmoto, Nightingale and den Nijs\cite{Thoul}
an explicit formula for the Hall conductance of 2D noninteracting electrons in
a periodic potential was derived. As it was found in Ref.\cite{Thoul}, such a
system has integral values of the Hall conductance in unit of $e^2/h$ if the
Fermi energy lies in a gap between magnetic subbands. This result has a
topological nature and as was mentioned in many works does not depend on the
detailed structure of the potential\cite{Dana}. In this paper we show that the
Hall conductance quantization law and the structure of the spectrum of magnetic
subbands can depend on the space symmetry and some potential parameters.

We study quantum Hall effect and quantum eigenstates in 2D electron system in
the presence of periodic potential which is formed by triangular antidots
located on the sites of the square lattice. In this system the inversion
center and four-fold rotation symmetry are absent. As a result, the quantum
states and Hall conductance quantization law differ from the case of the simple
square lattice potential. Indeed, in a zero magnetic field the electron energy
is invariant with respect to the changing of the quasimomentum sign. This
symmetry is a consequence of the invariance of Schr\"odinger equation about
the time reversal. In a magnetic field, the $t\to -t$ symmetry is violated
and, therefore, it is natural to assume that at the lack of inversion center
in the ${\bf r}$-space the energy of Bloch electron will be not an even
function of the ${\bf k}$, that is $E({\bf k})\ne E(-{\bf k})$. It may be
expected that in the last case a crystal possesses unusial physical properties.
At first time the problem under consideration was investigated in our previous
papers\cite{DPJL} for simplified model of periodic potential without the
inversion center. It was shown that the system without the inversion symmetry
in the ${\bf k}$-space has a non-trivial transport\cite{DPJL} and optical
properties. In particular, at the lack of inversion symmetry in the
${\bf k}$-space the so-called photogalvanic effect\cite{PG} must be observed,
i.e. in 2D electron gas the normally spreading electromagnetic wave causes
the lateral dc current.

It should be noted that the asymmetry of energy spectrum with respect to the
${\bf k}$ sign reversal can be observed also in systems with spin-orbit
interaction. This asymmetry is realized at the lack of inversion center of
the electric potential. In this case even in zero magnetic field, the electron
energy depends on the spin orientation, and only the Kramers degeneracy
$E({\bf k},1/2)=E(-{\bf k},-1/2)$ takes place. There are three
sources of this inversion asymmetry: a) bulk inversion asymmetry associated
with the common zinc-blend crystal structure\cite{DHaus}; b) asymmetry of
electric potential in heterojunctions\cite{Rash} and c) the asymmetry due
to different composition of both anions and cations in quantum well and
barrier\cite{STron}.

The Hall conductivity of the fully occupied magnetic subbands separated by
energy gaps was originaly calculated in the papers of
Refs.\cite{{Thoul},{Streda}} As it was shown later, the  quantization of
$\sigma_{xy}$ in this system has the topological nature. The topological
properties of the eigenfunctions corresponding to magnetic subbands at first
time was discussed by Novikov et al.\cite{Nov} The same problem was discussed
by Avron, Seiler, and Simon within the framework of homotopy theory\cite{Avron}.
Simon\cite{Simon} found the relation between the topological invariants and
the Berry geometric phase\cite{Berry}. Kohmoto showed that the quantized Hall
conductivity of a magnetic subband is determined by the topological
singularities of the wave function\cite{KohmNY}, and in terms of $e^2/h$
units this conductivity is equal to the first Chern number of the magnetic
subband taken with the opposite sign. However, Kohmoto gave no explicit
expressions for the calculations of conductivity. This problem was solved
by Usov\cite{UsJ}, who found interrelation between the Hall conductivity
$\sigma_{xy}$ of a subband and the wave function singularities (phase
branching points in the ${\bf k}$-space). As it was shown in Ref.\cite{DKh}
this method can be generalized also for the system with spin-orbital
interaction and, in particular, for the 2D hole gas in $p$-type
heterojunctions with lateral superlattice.

At the last decade, the structures with a 2D electron gas modulated by a surface
potential have been investigated experimentaly. The splitting
of magnetic levels into subbands in an $n$-type heterojunction
at the presence of a surface superlattice was observed for the first time
in Ref.\cite{Schl}. The effects of magnetic subbands formation have been
studied also in papers\cite{{KvK},{Gei}}, where the magnetotransport in
the structures with a surface superlattice was investigated.

It should be mentioned also the recent papers\cite{Lor} where the properties
of 2D electron gas in the field of lattices with triangular basis have been
investigated experimentally in the classical regime. In particular, Lorke with
co-authors studied the high- and low-frequency magnetotransport of a square
antidot lattice with a saw-tooth potential caused by the triangular antidots.
It was declared that under the far-infrared irradiation the broken symmetry of
the antidot array leads to a lateral photo-voltage. The authors of the
Ref.\cite{Lor} suppose that this effect is related to a new type of classical
skipping-orbit trajectories, which are absent in the commonly used circular
geometry.

Below we will demonstrate that in the presence of magnetic field
the asymmetry of electron energy spectrum in the ${\bf k}$-space takes place
for a spinless particles due to the lack of inversion cener of periodic
potential. The outline of this paper is as follows. Sec.~II presents the
description of the basic model and, in particular, of the geometry of
triangular antidots. In Sec.~III we calculate the Hall conductance of magnetic
subbands split off Landau levels due to the presence of periodic potential of
quantum antidots for the case when the number of magnetic flux quanta per unit
cell is equal to three. We demonstrate that with varying the shape of single
antidot only two different Hall conductance quantization laws $\sigma_{xy}(H)$
can be realized. The critical values of antidot geometric parameters at which
the change of quantization law takes place are determined. Finaly, the Hall
conductance which corresponds to two different quantization laws for the case
of three occupied magnetic subbands is calculated.

\section{Quantum states}
The Hamiltonian of an electron in uniform magnetic field and in two-dimensional
periodic potential has the form $\hat H=\hat H_0+V(x,y)$, where
$\hat H_0=({\bf\hat p}-e{\bf A}/c)^2/2m^{\ast}$ is the free electron
Hamiltonian in a uniform magnetic field. Here $c$ is the light velocity, $e$
and $m^{\ast}$ are the electron charge and effective mass, respectively. The
vector potential ${\bf A}$ of uniform magnetic field is chosen in the Landau
gauge ${\bf A}=(0,Hx,0)$, so that ${\bf H}\parallel z$.

The model periodic potential of triangular dot/antidot square lattice has the
form
$$
\displaylines{V(x,y)=V_0\bigg[\cos^2(\pi x/a)\cdot\cos^2(\pi y/a)+
\cos^2(\pi (x-d_x)/a)\cdot\cos^2(\pi (y-d_y)/a)+\cr
\hfill \cos^2(\pi (x+d_x)/a)\cdot\cos^2(\pi (y-d_y)/a)\bigg],\hfill\llap{(1)}\cr}
$$
where $V_0$ and $a$ are its amplitude and period. The potential (1) is periodic
with the period $a$ along the $x$ and $y$ directions and has three local
minima/maxima near the lattice sites. One can consider that on each site
the isosceles triangular dot/antidot is located. At $d_y=0$ and also when
$d_x=d_y=0$ or $a/2$ the symmetry of the lattice with respect to the inversion
is restored. At all other values of $d_x$ and $d_y$ the potential is
noncentrosymmetric. In Fig.1a the contourplots of the potential (1) are shown
for $d_y=5a/16$ and $d_x=1.1\, d_y$. One can see that in this case the lattice
potential has no inversion center and the square lattice is formed by
antidots with triangular basis. Four of the antidots are located at the
sites of the unit cell and one is disposed inside the cell. Thus, we deal with
the complex crystal lattice containing two atoms per unit cell. The
contourlines $V(x,y)=const$ at $d_x=0.05\, a$, $d_y=0.15\, a$ are shown
in Fig.1b. In sight, it is difficult to see the inversion symmetry here, and a
simple square lattice is formed by quantum antidots with quasi-cylindrical
shape.

The structure of electron quantum states is determined by the magnetic flux
through a unit cell \cite{Thoul}. If this flux (measured in flux quanta
$\Phi_0$) is a rational number $\Phi/\Phi_0=p/q=|e|Ha^2/2\pi\hbar c$, where
$p$ and $q$ are prime integers, the electron wave function which at the same
time is an eigenfunction of the magnetic translation operator\cite{LP} obeys
the Bloch-Peierls conditions
$$
\Psi_{\bf k}(x,y)=\Psi_{\bf k}(x+qa,y+a)\cdot {\rm e}^{-ik_xqa}\cdot
{\rm e}^{-ik_ya}\cdot {\rm e}^{-2\pi ipy/a}.\eqno(2)
$$
The vectors ${\bf a_m}=\{m_1qa,m_2a\}$ ($m_1,m_2$ are integer numbers)
define the so-called magnetic lattice of the crystal and, so, the magnetic
Brillouin zone (MBZ) is determined by the inequalities
$-\pi/qa\le k_x\le\pi/qa,\; -\pi/a\le k_y\le\pi/a$.

The periodic potential leads to the mixing and splitting of Landau levels
into magnetic subbands. But if the inequalities
$$
\hbar\omega_c\gg V_0;\quad l_H\ll a\eqno(3)
$$
are fulfilled (here $l_H$ and $\omega_c$ are the magnetic length and cyclotron
frequency, respectively), one can neglect the interaction between different
Landau levels. Note that the conditions (3) can be easily realized
experimentally in artificial surface superlattices. So, at $a=80\, nm$,
$p/q=3$ and $V_0\approx 1\; meV$, the magnitude of the magnetic field must be
equal to $H\simeq 2\cdot 10^4\; Oe$.

The electron wave function of the $\mu$th magnetic subband which satisfies
to the conditions (2) can be decomposed on oscillator functions $\varphi_N$
of the $N$th Landau level\cite{Thoul}

$$
\displaylines{\Psi_{{\bf k},\mu}^N(x,y)=\sum\limits_{n=1}^{p}C_{n\mu}^N({\bf k})
\sum\limits_{\ell=-\infty}^{+\infty}\varphi_{N}\bigg[\frac{x-x_0-\ell qa-
nqa/p}{l_H}\bigg]\times\cr
\hfill \times{\rm e}^{ik_yy}{\rm e}^{ik_x(\ell qa+nqa/p)}
{\rm e}^{2\pi iy(\ell p+n)/a},\hfill\llap{(4)}\cr}
$$
where $x_0=c\hbar k_y/|e|H=k_y l_H^2$. Let us define also the functions
$$
u_{{\bf k},\mu}^N(x,y)=\Psi_{{\bf k},\mu}^N(x,y)\exp(-ik_x x)\exp(-ik_y y),
$$
which satisfy to the following conditions
$$
u_{{\bf k},\mu}^N(x+qa,y)\exp(-2\pi ipy/a)=u_{{\bf k},\mu}^N(x,y+a)=
u_{{\bf k},\mu}^N(x,y)
$$
and are the eigenfunctions of the Hamiltonian
$$
\hat H({\bf k})=\frac{1}{2m^{\ast}}(-i\hbar\partial_x+\hbar k_x)^2+
\frac{1}{2m^{\ast}}(-i\hbar\partial_y+\hbar k_y-eHx/c)^2+V(x,y).
$$

In the representation of symmetrized linear combinations of Landau functions
the stationary Schr\"odinger equation $\hat H\Psi=E\Psi$ has the form
$$
H_{nm}^N({\bf k})C_{m\mu}^N({\bf k})=[E_N^0(p/q)\delta_{nm}+
V_{nm}^N({p/q,\bf k})]C_{m\mu}^N=E_{\mu}^N({\bf k})C_{n\mu}^N({\bf k}),
\eqno(5)
$$
where $E_N^0=\hbar\omega_c(N+1/2)$. Here, the eigenvector components
obey the relations $C_{n+p,\mu}^N({\bf k})=C_{n\mu}^N({\bf k})$.

The hermitian $p\times p$ matrix $V_{nm}^N$ of the potential (1) has the
quasi-three-diagonal form where the diagonal elements are equal to
$$
D_n=\frac{3V_0}{4}+A_x\frac{V_0}{4}\cos\bigg(2\pi n\frac{q}{p}+
\frac{qk_ya}{p}\bigg)\exp\bigg(-\frac{\pi q}{2p}\bigg)
L_N^0\bigg(\frac{\pi q}{p}\bigg),\eqno(6.a)
$$
and upper off-diagonal elements are
$$
\displaylines{M_n=\frac{V_0}{4}\Bigg[\frac{A_y}{2}+iS_y\Bigg]
\exp\bigg(i\frac{k_xqa}{p}\bigg)\exp\bigg(-\frac{\pi q}{2p}\bigg)
L_N^0\bigg(\frac{\pi q}{p}\bigg)+\cr
\hfill \frac{V_0}{4}\Bigg[\frac{B}{2}+iC_xS_y\Bigg]
\exp\bigg(i\frac{k_xqa}{p}\bigg)\cos\bigg(2\pi (n+1/2)\frac{q}{p}+
\frac{qk_ya}{p}\bigg)\times\hfill\cr
\hfill \exp\bigg(-\frac{\pi q}{p}\bigg)
L_N^0\bigg(\frac{2\pi q}{p}\bigg).\hfill\llap{(6.b)}\cr}
$$
Here $L_N^0(\xi)$ are Laguerre polynomials. Moreover, the elements
$V_{p,1}=V_{1,p}^{\ast}$ are equal to $M_p$. In Eqs.(6) the parameters
$A_x,\, A_y,\, B,\, C_x$ and $S_y$ are defined by following relations:
$$
\displaylines{C_x=\cos(2\pi d_x/a),\;\; C_y=\cos(2\pi d_y/a),\;\;
S_y=\sin(2\pi d_y/a),\cr
\hfill A_x=1+2C_x,\;\; A_y=1+2C_y,\;\; B=1+2C_xC_y. \hfill\cr}
$$

As an example let us calculate analyticaly the Bloch electron states for the
case when the value of magnetic flux quanta number $p/q=3/1$. The energy
spectrum in this case was found by solving a cubic characteristic equation of
the system of homogenious linear equations (5). At fixed ${\bf k}$ there are
$p$ different roots of the secular equation of the system (5). Thus, at the
rational number of magnetic flux quanta per unit cell the periodic potential
leads to the splitting of each Landau level into $p$ magnetic subbands. As an
example, the contourplots of electron energy $E_1^1({\bf k})$ in the lowest
subband of the first Landau level $N=1$ are shown in Fig.2a. This spectrum is
plotted for the potential (1) at the same parameters as for Fig.1a. Here, one
can see the absence of inversion symmetry of the function $E({\bf k})$ in the
${\bf k}$-space. But, the spectrum remains symmetric under the replacement of
the $k_y$ sign. The last feature of the dispersion law is due to the symmetry
of the potential $V(x,y)$ with respect to the changing of the sign of $x$
coordinate, that is $x\to -x$. The contourlines of the function
$E_1^0({\bf k})$ for the lowest magnetic subband of the zeroth Landau level
$N=0$ are shown in Fig.2b. Here the parameters $d_x$ and $d_y$ are the same as
in Fig.1b, and the violation of inversion symmetry of the spectrum here is not
so remarkable as in Fig.2a.

The contourplots of the probability density $|\psi (x,y)|^2$ at ${\bf k}=0$,
$p/q=3/1$ for the lowest magnetic subband $E_1^0({\bf k})$ of zero Landau level
$N=0$ are shown in Fig.3. The lattice parameters $d_x$ and $d_y$ in this
case are the same as for Fig.1a and Fig.2a. One can see that the symmetry of
the probability density for the ground state corresponds to the symmetry of
the potential. Note that the probability density is equal to zero in three
points of its minima. In accordance with general theory\cite{KohmNY}, the
clockwise tracing of these points lying in the magnetic unit cell gives the
total phase incursion of the wave function equal to $-2\pi p$. It could be
mentioned also that under the $V_0$ sign reversal which means the transition
from antidot to dot lattice the changing of numbers and positions of the wave
function zeros occur. However, the total phase incursion remains the same.

\section {Hall conductance}
As was mentioned above, the quantum Hall effect in 2D periodic potential has a
topological nature\cite{{Thoul},{UsJ}}. We have investigated the influence of
the shape of triangular antidots assembled into lattice on topological
invariants (first Chern numbers) of magnetic subbands, which determine their
Hall conductance.

The coefficients $C_{n\mu}^N({\bf k})$ can be chosen proportional to the
algebraic adjunct $D_{jn}({\bf k})$ of any (for example, $j$th) row of the
secular determinant of the system of Eqs.(5) at $E=E_{\mu}({\bf k})$. According
to Ref.\cite{UsJ}, the components of a normalized eigenvector can be presented in
the following form
$$
C_{n\mu}^{N}({\bf k})=R_{n\mu}^{N}({\bf k})\exp(i\phi_{n\mu}^{N}({\bf k})),\,
R_{n\mu}^{N}({\bf k})=\bigg(D_{nn}({\bf k})/\sum_{s=1}^p
D_{ss}({\bf k})\bigg)^{1/2},\; n\ne j.\eqno(7)
$$
In Eqs.(7), $D_{nn}({\bf k})$ is the algebraic adjunct to the matrix element
$[H_{nn}^N-E_{\mu}^N]$, and the phase $\phi_{n\mu}^{N}({\bf k})$ is determined
by the relations
$$
\cos\phi_{n\mu}^{N}=\frac{{\bf Re}\, D_{jn}}{|D_{jn}|},\;
\sin\phi_{n\mu}^{N}=\frac{{\bf Im}\, D_{jn}}{|D_{jn}|},\; n\ne j.\eqno(8)
$$

It can be shown that the component $C_{j\mu}^{N}({\bf k})$ is a purely real
function, and it vanishes at some special points ${\bf k}_m$ of the extended
magnetic Brillouin zone
$-\pi/qa\le k_x\le\pi/qa,\;-\pi p/qa\le k_y\le \pi p/qa$ introduced by
Usov\cite{UsJ}. The other components of the eigenvector are complex functions
of real vector variable ${\bf k}$ and in singular points ${\bf k}_m$ have
no definite limit. Thus, such singularities are phase branching points in
${\bf k}$-space for coefficients $C_{n\mu}^{N}({\bf k})$ at $n\ne j$. Under
tracing the points ${\bf k}_m$ in negative (clockwise) direction each of the
components $C_{n\mu}^{N}({\bf k})$ ($n\ne j$) gets the phase incursion
$2\pi S({\bf k}_m)$, where $S({\bf k}_m)$ are integer numbers. The total phase
incursion obtained by tracing all of the singularities ${\bf k}_m$ of any
coefficient $C_{n\ne j,\mu}^{N}({\bf k})$ determines the topological invariant
(first Chern number) of a chosen subband.

Let us define the Hall conductance as\cite{Thoul}
$$
\sigma_{xy}=\frac{ie^2}{2\pi h}\sum\limits_{\mu}\int {d^2k}
\int {d^2r}\Bigg(\frac{\partial u_{{\bf k},\mu}^N}{\partial k_x}
\frac{\partial u_{{\bf k},\mu}^{N\ast}}{\partial k_y}-c.c.\Bigg),\eqno(9)
$$
where the sum on $\mu$ and integrations are over the occupied magnetic subbands
and over the unit cells in ${\bf k}$- and ${\bf r}$-space, respectively. As was
shown by Usov\cite{UsJ}, the Hall conductance of the $\mu$th subband may be
presented as a sum of two terms
$\sigma_{xy}=\sigma_{xy}^{(1)}+\sigma_{xy}^{(2)}$, where
$$
\sigma_{xy}^{(1)}=\frac{e^2}{ph}\quad {\rm and}\quad
\sigma_{xy}^{(2)}=-i\frac{e^2}{2\pi h}\oint\limits_{\gamma}
\sum\limits_{n=1}^p J_n^{\alpha}\, dk_{\alpha}.\eqno(10)
$$
Here, the vector ${\bf J}_n$ is determined by its components
$$
J_n^{\alpha}=\frac{1}{2}\Bigg[C_{n\mu}^{N\ast}
\frac{\partial C_{n\mu}^N}{\partial k_{\alpha}}-c.c.\Bigg],\quad \alpha=x,y.
\eqno(11)
$$
In accordance with Eqs.(7) and (8) one can obtain that
$$
J_n^{\alpha}=iR_n^2\frac{\partial\varphi_n}{\partial k_{\alpha}}.\eqno(12)
$$
The counter-clockwise path integral in Eq.(9) is over the boundaries of the MBZ
and over the contours which arround all of the singularities ${\bf k}_m$ of any
coefficient $C_{n\ne j,\mu}^{N}({\bf k})$. But, as was shown in \cite{UsJ} the
contour of integration in Eq.(9) may be chosen by the same way in the extended
MBZ which has the size in $k_y$ direction larger than the original MBZ in
$p/q$ times. Thus, the contribution to the integral over the boundaries of the
extended MBZ becomes zero, and the Hall conductance is determined by the
singularities ${\bf k}_m$ lying in the extended MBZ. Finaly, the explicit
expression for Hall conductance of the $\mu$th magnetic subband corresponding
to the $N$th Landau level takes the form\cite{UsJ}
$$
\sigma_{xy}=-\frac{e^2}{h}\Bigg[\frac{1}{p}+\frac{q}{p}\sum\limits_{{\bf k}_m}
S({\bf k}_m)\Bigg].\eqno(13)
$$
Note that the integer numbers $S({\bf k}_m)$ determine the phase incursion of
any of the coefficients $C_{n\mu}^{N}({\bf k})$ at $n\ne j$.

Our calculations have shown that in the framework of our model two different
quantization laws for the Hall conductance can be realized. The corresponding
results are displayed in Fig.4. Here, the parameters $d_x$ and $d_y$ vary in
the one-fourth part of the square unit cell. The Hall conductance of the fully
occupied lowest magnetic subband $E_1^0({\bf k})$ of the zeroth level at
$p/q=3/1$ takes different values in two regions (I) and (II) shown in Fig.4.
In the region (I) which is painted in gray colour the Hall conductance of the
subband $E_1^0({\bf k})$ is equal to $-e^2/h$. In the area (II) the Hall
conductance $\sigma_{xy}=0$. It was found that under the crossing the
separating line in Fig.4 the lowest $E_1^0({\bf k})$ and the middle
$E_2^0({\bf k})$ magnetic subbands of zero Landau level touch each other at
the certain point ${\bf k}_0$ in MBZ. At the same time the topological
invariants of these subbands are changing instantly.
So, in the area (II) shown in Fig.4 the Hall conductance of the lowest subband
$E_1^0({\bf k})$ becomes zero, but the conductance of the middle one is equal
to $-e^2/h$. In particular, as one can see from Fig.4 for $d_x,\, d_y\ll a$
the conductance of magnetic subbands is the same as for the case of
centrosymmetric potential\cite{DPJL}. Moreover, the quantization law which is
typical for the lattices with inversion symmetry is realized also even for
$d_x$ or $d_y$ approximately equal to $a/2$. Thus, when $d_x=a/2$ and $d_y=0$
or $a/2$ the spatial inversion symmerty is reconstructed. In this case each
unit cell of the lattice contains two different atoms and Hall conductance of
the lowest subband is the same as for centrosymmetric lattice, that is,
$-e^2/h$. It was found also that the conductance of upper magnetic subband
does not depend on antidot geometry and equals zero for the magnetic flux
$p/q=3/1$.

The transformation of Hall conductance quantization rule for magnetic
subbands of zero Landau level under the crossing the separating line between
(I) and (II) regions is illustrated in Fig.5. Along the horizontal axis the
position of Fermi energy level with respect to the location of magnetic
subband are displayed. One can see that there are two different quantization
laws of Hall conductance depending on the values of the parameters $d_x$ and
$d_y$.

Since the difference between these quantization laws depends on the lattice
geometry, it can be observed experimentally on different triangular lattices.
Note, that for the lattice with the period $a=80\, nm$, $p/q=3$ and
$V_0\approx 1\; meV$, the width of the magnetic subbands have to be smaller
than the collision broadening of energy levels.

It should be mentioned that for the other values of the flux number $p/q$
the changing of the topological invariants of magnetic subbands occurs at
the moment of their touching. For example, it was found\cite{PDem} that
for some critical values of $d_x$ and $d_y$ at $p/q=4/1$ the energy gap
between the second and the third magnetic subbands of the zeroth Landau
level disappears. At the same time the Hall conductances of these subbands
undergo a sudden changes.

\section{Summary}
In conclusion, we have studied the quantum states of 2D electrons moving
in the periodic potential of square lattice with triangular antidots in its
sights. In this system the electron energy is not an even function of
quasimomentum defined in the MBZ. The singular points of the wave functions
in ${\bf r}$- and ${\bf k}$-spaces having the topological nature are found
for different geometrical parameters of antidots. The topological invariants
of magnetic subbands (first Chern numbers) which define their Hall conductances
depend drastically on the symmetry of the lattice potential. It was shown that
by changing the single antidot parameters two different Hall conductance
quantization laws $\sigma_{xy}(H)$ can be observed. The boundaries of the
areas of antidot parameters $d_x$ and $d_y$ at which the Hall conductance
quantization law changes are found. The desired values of the system parameters
at which the experimental observation of the different Hall conductance
quantization laws is possible are indicated.

\section*{Acknowledgments}
This work has been supported by the Russian Foundation for Basic Researches
(Grant No. 03-02-17054), the Ministry of Education of Russian Federation
(Grant No. PD02-1.2-147), and by the Program "Universities of Russia".

\end{document}